\documentclass[aps,amssymb,amsmath,pra,twocolumn,showpacs,superscriptaddress,longbibliography]{revtex4-2}

\usepackage{ulem}
\usepackage{comment} 

\usepackage{graphicx}
\graphicspath{{figs_main/}{figs_SI/}}

\usepackage{dcolumn}
\usepackage{bm}
\usepackage{subfigure}
\usepackage{color}
\usepackage{soul}

\usepackage{amsfonts,verbatim}

\usepackage[margin=0.6in]{geometry}

\usepackage{tikz}

\usepackage[english]{babel}
\usepackage{dsfont}
\usepackage{latexsym}
\usepackage{float}
\usepackage{afterpage}
\usepackage{enumitem}
\usepackage{mleftright}

\definecolor{lR}{rgb}{1, 0.8, 0.79}

\usepackage{eso-pic, graphicx}

\usepackage{listings}
\usepackage{multirow}
\usepackage{xcolor,colortbl}

\usepackage{bbm}
\usepackage{upgreek}
\usepackage{newtxtext,newtxmath}

\newcommand{\ignore}[1]{}

\newcommand{\nocontentsline}[3]{}
\newcommand{\tocless}[2]{\bgroup\let\addcontentsline=\nocontentsline#1{#2}\egroup}

\definecolor{Ablue}{rgb}{0.96,0.24,0.00}

\definecolor{Abluetitle}{rgb}{0.,0.24,0.51}

\definecolor{orange}{rgb}{0.96,0.24,0.00}

\definecolor{darkred}{rgb}{0.55, 0.0, 0.0}

\definecolor{darksalmon}{rgb}{0.91, 0.59, 0.48}
\definecolor{maroon}{cmyk}{0,0.87,0.68,0.32}

\setcitestyle{numbers,square,citesep={,\kern-.24em}}

\definecolor{mustard}{rgb}{1.0, 0.86, 0.35}

\addto\captionsenglish{\renewcommand{\figurename}{Fig.}}

\definecolor{Gray}{gray}{0.85}
\definecolor{LightCyan}{rgb}{0.88,1,1}
\newcolumntype{a}{$>${\columncolor{Gray}}c}
\newcolumntype{b}{$>${\columncolor{White}}c}

\usepackage{array}
\newcolumntype{L}[1]{$>${\raggedright\let\newline\\\arraybackslash\hspace{0pt}}m{#1}}
\newcolumntype{C}[1]{$>${\centering\let\newline\\\arraybackslash\hspace{0pt}}m{#1}}
\newcolumntype{R}[1]{$>${\raggedleft\let\newline\\\arraybackslash\hspace{0pt}}m{#1}}

\newcolumntype{P}[1]{>{\centering\arraybackslash}p{#1}}
\newcolumntype{M}[1]{>{\centering\arraybackslash}m{#1}}

\usepackage[colorlinks=true , citecolor=black,urlcolor=blue]{hyperref}

\newcommand{\xg}{\gamma}
\newcommand{\xt}{\vartheta}

\newcommand{\xr}{\rho}

\newcommand{\app}{\approx}

\newcommand{\Cs}{{}^{13}\R{C}}

\newcommand{\Hs}{{}^{1}\R{H}}

\newcommand{\mw}{\R{MW}}

\newcommand{\mH}[0]{\mathcal{H}}

\newcommand{\rt}{\rightarrow}

\newcommand{\beq}{\begin{equation}}
\newcommand{\eeq}{\end{equation}}
                  
\newcommand{\benum}{\begin{enumerate}}
\newcommand{\eenum}{\end{enumerate}}
                    
\newcommand{\bit}{\begin{itemize}}
\newcommand{\eit}{\end{itemize}}
\newcommand{\xhat}{\hat{\T{x}}}
\newcommand{\yhat}{\hat{\T{y}}}
\newcommand{\zhat}{\hat{\T{z}}}
\newcommand{\ahat}{\hat{\T{a}}}
\newcommand{\bhat}{\hat{\T{b}}}
\newcommand{\chat}{\hat{\T{c}}}

\newcommand{\bea}{\begin{eqnarray}}
\newcommand{\eea}{\end{eqnarray}}

\newcommand{\bev}{\begin{verbatim}}
\newcommand{\eev}{\end{verbatim}}

\newcommand{\qt}{\tau}

\newcommand{\T}[1]{\textbf{#1}}
\newcommand{\I}[1]{\textit{#1}}
\newcommand{\R}[1]{\textrm{#1}}

\newcommand{\zfl}[1]{\protect\label{fig:#1}}
\newcommand{\zfr}[1]{\figurename\,\ref{fig:#1}}

\newcommand{\ket}[1]{\left\vert{#1}\right\rangle}

\newcommand{\ba}{\left\{ \begin{array}{lr}}
\newcommand{\ea}{\end{array}\right.}

\newcommand{\blist}[1]{
 \begin{list}{#1}
 \begin{align}
	 arrow
 \end{align}
 $\checkmark\star
  { \setlength{\itemsep}{3pt}
     \setlength{\parsep}{2pt}
     \setlength{\topsep}{3pt}
     \setlength{\partopsep}{0pt}
     \setlength{\leftmargin}{1em}
     \setlength{\labelwidth}{1em}
     \setlength{\labelsep}{0.5em} } }
\newcommand{\elist}{
  \end{list}  }

\DeclareMathSymbol{\vartheta}{\mathalpha}{letters}{"12}
\DeclareMathSymbol{\theta}{\mathalpha}{letters}{"23}
\DeclareMathSymbol{\phi}{\mathalpha}{letters}{"27}
\DeclareMathSymbol{\varphi}{\mathalpha}{letters}{"1E}

\newcommand{\bef}
{
\begin{figure}[htbp]
\centering
}

\newcommand{\eef}{\end{figure}}

\mleftright
\makeatletter
\@addtoreset{section}{part}
\makeatother

\newcommand{\affA}{Department of Chemistry, University of California, Berkeley, Berkeley, CA 94720, USA.}
\newcommand{\affB}{Department of Chemical and Biomolecular Engineering, University of California, Berkeley,
CA 94720, USA.}
\newcommand{\affC}{Materials Science Division, Lawrence Berkeley National Laboratory, Berkeley, CA 94720.}
\newcommand{\affD}{Department of Physics \& Astronomy, University College London, Gower Street, London, WC1E 6BT, UK.}
\newcommand{\affE}{Chemical Sciences Division,  Lawrence Berkeley National Laboratory,  Berkeley, CA 94720, USA.}
\newcommand{\affF}{CIFAR Azrieli Global Scholars Program, 661 University Ave, Toronto, ON M5G 1M1, Canada.}
\newcommand{\affG}{Department of Pure and Applied Science, University of Urbino Carlo Bo, Urbino, I-61029, Italy.} 
\newcommand{\affI}{Department of Physics, Guru Nanak Dev University, Amritsar, Punjab 143005, India.}

\begin{document}
\title{Room-temperature quantum sensing with photoexcited triplet electrons in organic crystals}
\author{Harpreet Singh}\affiliation{\affA}\affiliation{\affI}
\author{Noella D'Souza}\affiliation{\affA}\affiliation{\affE}
\author{Keyuan Zhong}\affiliation{\affA}
\author{Emanuel Druga}\affiliation{\affA}
\author{Julianne Oshiro}\affiliation{\affB}
\author{Brian Blankenship}\affiliation{\affA}
\author{Riccardo Montis}\affiliation{\affG}
\author{Jeffrey A. Reimer}\affiliation{\affB}\affiliation{\affC}
\author{Jonathan D. Breeze}\affiliation{\affD}
\author{Ashok Ajoy}\email{ashokaj@berkeley.edu}\affiliation{\affA}\affiliation{\affE}\affiliation{\affF}

\begin{abstract}
Quantum sensors have notably advanced high-sensitivity magnetic field detection. Here, we report quantum sensors constructed from polarized spin-triplet electrons in photoexcited organic chromophores, specifically focusing on pentacene-doped para-terphenyl (${\app}$0.1\%). We demonstrate essential quantum sensing properties at room temperature (RT): optically-generated electronic polarization and state-dependent fluorescence contrast, by leveraging differential pumping and relaxation rates between triplet and ground states. We measure high optically-detected magnetic resonance (ODMR) contrast ${\app}16.8\%$ of the triplet states at RT, along with long coherence times under spin echo and CPMG sequences, $T_2{=}2.7\mu$s and $T_2^{\R{DD}}{=}18.4\mu$s, respectively, limited only by the triplet lifetimes. The material offers several advantages for quantum sensing, including the ability to grow large ($\I{cm}$-scale) crystals at low cost, absence of paramagnetic impurities, and electronic diamagnetism when not optically illuminated. Utilizing pentacene as a representative of a broader class of spin triplet- polarizable organic molecules, this study highlights new potential for quantum sensing in chemical systems.
\end{abstract}

\maketitle
\pagebreak

\section{Introduction}
Quantum sensors~\cite{Degen17} \I{(q-sensors)} have substantially expanded the scope for ultra-sensitive detection of magnetic fields and related parameters, e.g. temperature~\cite{neumann2013high,doherty2014temperature}, pressure~\cite{doherty2014electronic}, and rotation~\cite{Ajoy12g, Ledbetter12, Jaskula19, Jarmola21}. Traditional q-sensors have predominantly employed semiconductor defect color centers, typified by diamond NV centers~\cite{doherty2012theory}. They host highly coherent electronic spins that can be optically initialized and read out at room temperature (RT). Single~\cite{Taylor08} or densely packed NV center ensembles~\cite{Pham11} facilitate magnetometry over diverse length scales, even enabling the construction of compact, deployable magnetometry devices~\cite{Wolf15}.

Recently, there has been renewed interest in q-sensors constructed from chemical systems, including molecules~\cite{maylander2021exploring} or rare-earth ions~\cite{bayliss2020optically,serrano2022ultra}. The versatility of synthetic chemistry enables fabrication of tunable, atomically-defined sensor assemblies~\cite{zadrozny2017porous} at large scales. However, a critical gap remains in determining chemical systems that support coherent electronic control, optical initialization, and readout at RT, all features readily accessible with NV centers.

To address this challenge we introduce q-sensors based on photoexcited spin-triplet electrons in organic chromophore crystals. We employ pentacene, (\zfr{mfig1}A), a historically-significant molecule for single-molecule fluorescence~\cite{Moerner89}, EPR~\cite{Wrachtrup93,kohler_magnetic_1993,Sloop81,Yu84,Yang00} and room-temperature MASERs~\cite{Oxborrow12, breeze2017}, doped into a para ($\I{p}$)-terphenyl (PDP) host matrix as a model system. However, the approach here is readily applicable to a broad range of adjacent systems, including other acenes and porphyrins~\cite{tait2015triplet,richert2017constructive}. We demonstrate essential quantum sensing properties — optical initialization and readout, bright optical transitions, and coherent microwave (MW) control — under ambient conditions. This exploits differential pumping rates into the triplet state, following intersystem crossing (ISC), for initialization and varied relaxation rates of the triplet sub-levels back to the singlet ground state to produce state-dependent fluorescence contrast for readout (\zfr{mfig1}E-F). We use this to demonstrate optically detected magnetic resonance (ODMR) of spin transitions, with high contrast (16.8\%), and long coherence times ($T_2^{\R{DD}}{=}18.4 \mu$s), facilitating coherent control of the molecular electronic spin at RT.

\begin{figure*}[t]
  \centering
  {\includegraphics[width=0.97\textwidth]{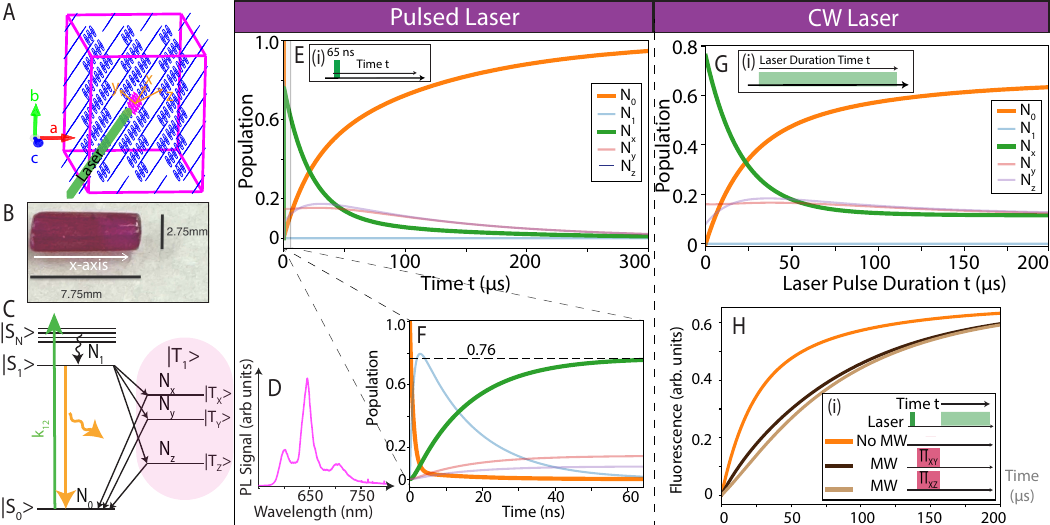}}
 
  \caption{\T{System and Principle.} (A) \I{Crystal structure} of pentacene molecule doped in $\I{p}$-terphenyl (PDP). Primitive lattice vectors are marked $\ahat$, $\bhat$, and $\chat$, with lattice parameters: $a{=} 24 $\AA, $b{=}28$ \r{A}, $c{=} 41$ \r{A}~\cite{prentice-jctc-22,rietveld1970x}. Molecular axis are marked $\xhat$, $\yhat$, and $\zhat$. (B) \I{Photograph} of representative crystal. (C) \I{Energy level structure:} with marked singlet ground ($\ket{S_0}$) and excited states ($\ket{S_1}$, $\ket{S_N}$) and photoexcited triplet state $\ket{T_1}$. (D) \I{PL spectrum} at RT taken with a 600 nm long pass filter. (E-F) \I{Initialization under pulsed laser}. Simulation of triplet sublevel populations after a 65ns 532nm pulse (\I{Inset} (i)) using a pumping rate of $k_{12}{=}10^9$s$^{-1}$. Ground singlet $\ket{S_0}$ and $\ket{T_x}$ populations ($N_0$ and $N_x$) are bolded. (F) Zoom in to a ${\app}60$ns window shows transient polarization of ${\app}$76\% developing in $\ket{T_x}$ due to differential (ISC) rates into the triplet state. (G) \I{Action of cw-laser} for polarized initial state ($N_x{=}0.76$) in (F) using a pumping rate of $k_{12}{=}10^4$ s$^{-1}$. (H) \I{Tracked population of $\ket{S_0}$ state}, reflective of the fluorescence signal, shown by the orange trace for case in (G). Dark brown and light brown traces show $\ket{N_0}$ where a resonant MW $\pi$-pulse is applied to the $T_{xy}$ or $T_{xz}$ transitions respectively. Transient spin-dependent fluorescence emerges, producing ODMR contrast.
}
\zfl{mfig1}
\end{figure*}

\section{System and Principle}
$\I{cm}$-size PDP single crystals (\zfr{mfig1}B) were grown using the Bridgman-Stockbarger technique~\cite{Stockbarger-rsI-36,bridgman1964certain}. Measurements were conducted on a ${\sim}$0.1\% doped crystal (\zfr{mfig1}B), focusing on a ${\sim}5700 \mu$m$^3$ spatial volume hosting ${\sim}10^{11}$ molecules. Molecular coordinate axes are depicted in \zfr{mfig1}A, where $\xhat$ denotes the molecular long-axis (\zfr{mfig1}A); with $\yhat$ and $\zhat$ transverse to it, the latter being out-of-plane.

\zfr{mfig1}C illustrates the $\pi$-electron energy level structure, consisting of a ground singlet state $\vert S_0\rangle$ and excited singlet $\vert S_1\rangle$ of lifetime ${\app}$22 ns~\cite{kohler1999magnetic}, separated by a ${\app}$2 eV gap and a metastable triplet state $\vert T_1\rangle$, with sublevels labelled $\ket{T_{x,y,z}}$. Optical excitation at 532 nm leads to fluorescence emission, and phonon-mediated ISC into an excited triplet state $\ket{T_2}$ (not shown for simplicity) with high quantum efficiency ${\app}62.5\%$~\cite{Takeda02} . The $\ket{T_2}$ electrons rapidly undergo internal conversion (IC) into the $\ket{T_1}$ triplet state via spin-conserving, non-radiative mechanisms~\cite{Oxborrow12}. \zfr{mfig1}D displays the RT photoluminescence (PL) spectrum under a 600nm long-pass filter; unlike cryogenic measurements~\cite{Patterson84}, the zero-phonon line is not visible and phonon sidebands dominate the PL spectrum. The photoexcited triplet spin Hamiltonian is described by ${\cal H}_{\R{sys}}{=}D\left(S_{z}^{2}-\frac{2}{3}\right)+E(S_{x}^{2}-S_{y}^{2})$, where $\boldsymbol{S}$ is a spin-1 Pauli operator, and $D {\approx}1392$ MHz and $E{\approx}-53$ MHz are the zero-field (ZF) splitting parameters~\cite{Yang00}.

\zfr{mfig1}E simulates the development of triplet electronic spin polarization under pulsed laser illumination of duration $t_p{=}65$ ns, assuming an optical pumping rate $k_{12}{=} 10^9$ s$^{-1}$, and employing previously-measured population and decay rates~\cite{Wu19, wu2020room}. $N_{x,y,z}$ refer to transient populations in the $\ket{T_{x,y,z}}$ triplet states, while $N_{0,1}$ refer to those of the  $\ket{S_{0,1}}$ singlet levels, respectively. Polarization occurs as a result of the differential ISC rates from $\ket{S_1}{\rt}\ket{T_{x,y,z}}$. This is evident in \zfr{mfig1}F. In the 80 ns after the pulse, $N_x{\app}0.76$ triplet polarization is achieved. At longer times, the spin population returns to $\ket{S_0}$ (\zfr{mfig1}E).

$\ket{T_1}$ population levels are readout by optical fluorescence measurement. \zfr{mfig1}G shows the evolution of triplet state sub-level populations under weak $\I{cw}$-532nm illumination, after polarization as in \zfr{mfig1}F. $k_{12}$=$10^4$ s$^{-1}$ is assumed. The fluorescence emission rate is proportional to singlet ground-state population, $N_0$, (orange line in \zfr{mfig1}H). Upon application of a MW $\pi$-pulse resonant with a triplet transition, the transient fluorescence profile changes due to differential relaxation rates from $\ket{T_1}{\rt}\ket{S_0}$, depending on the triplet sub-levels coupled by the MW field. Brown lines in \zfr{mfig1}H show this for $T_{xy}$ $(\vert T_x\rangle {\leftrightarrow} \vert T_y\rangle$) and $T_{xz}$ transitions. Different fluorescence emission rates forms the basis for obtaining spin-state dependent fluorescence contrast. 

\begin{figure*}[t]
  \centering
  {\includegraphics[width=0.97\textwidth]{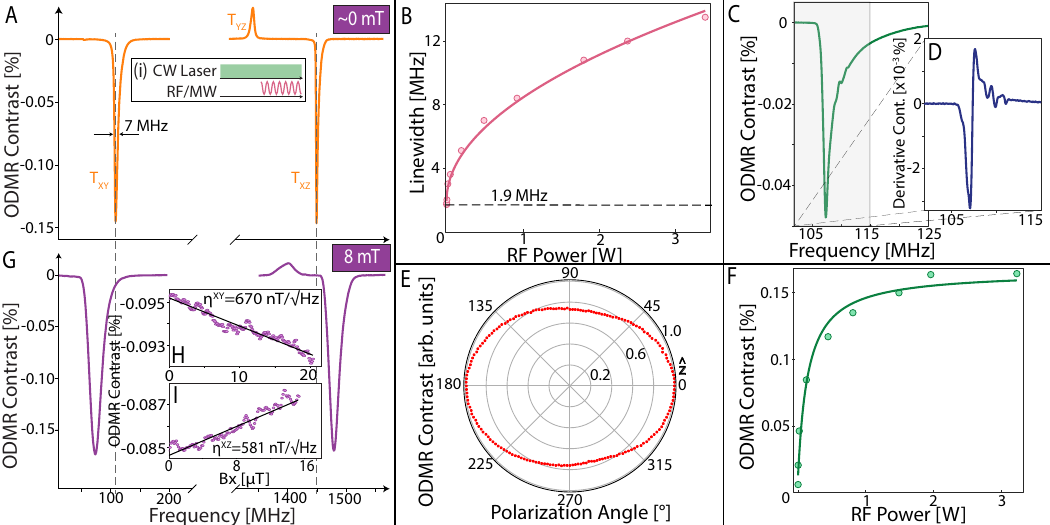}}
  \caption{\T{ODMR of the photoexcited triplet.} (A) \I{ODMR spectrum} at zero-field measured using cw-illumination protocol (\I{Inset} (i)) at 0.6 W MW power. Triplet transitions occur at 108, 1340, and 1448 MHz; linewidth $\ell{=}7$MHz. $T_{yz}$ transition contrast is inverted in sign.  
  (B) \I{Linewidth dependence} on power for $T_{xy}$ transition, yielding a power unbroadened linewidth $\ell_0{\app}$1.9 MHz.
  (C) \I{Zoom in} to the $T_{xy}$ transition at low power (1.4mW) shows asymmetric ODMR lineshape. 
  (D) \I{Lineshape derivative} of (C) shows spectral features from hyperfine coupling to $\Hs$ nuclei.
  (E) \I{Excitation beam polarization} of ODMR contrast, showing ${\app}27\%$ variation. 
  (F) \I{Contrast dependence} on power for the $T_{xy}$ transition. Solid line is a fit.
  (G) ODMR spectrum at $B_0{=}8$mT along $\xhat$. Dashed vertical lines highlight peak shifts relative to the zero-field spectrum. \I{DC-field sensitivities} for (H) $T_{xy}$ and (I) $T_{xz}$ with bias field $B_0$ along $\xhat$. \I{Points:} experimental data estimated from right spectral edges, and error bars reflecting linewidths. \I{Solid lines:} simulations. 
}
\zfl{mfig2}
\end{figure*}

\begin{figure}
  \centering
  {\includegraphics[width=0.49\textwidth]{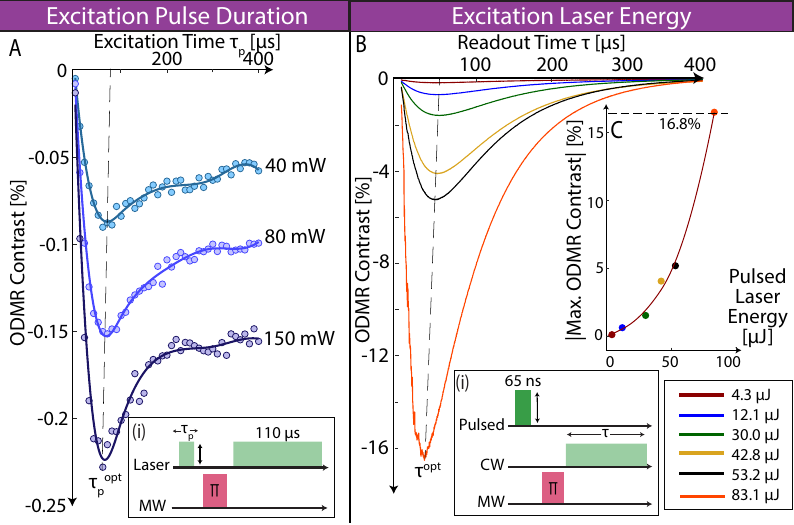}}
  \caption{\T{ODMR under pulsed illumination} for $T_{xy}$ transition. (A) ODMR profiles for varying laser pulse length $\qt_p$. Fluorescence is measured during the second 110$\mu$s $\I{cw}$-laser pulse (\I{Inset} (i)). Dashed line shows $\qt_p^{\R{opt}}$, which decreases with increasing power and increased contrast. (B) ODMR profiles with pulsed laser initialization (65ns), and varying periods $\qt$ of fluorescence measurement under 150mW $\I{cw}$-illumination (\I{Inset} (i)). Slanted dashed line show peaks shift with increasing pulsed laser energy. (C) Maximum absolute ODMR contrast with pulsed laser energy reaches ${\app}$16.8\% at 83mJ. Theoretically expected maximum contrast is 58\% (see SI). Dashed line is a guide to eye.}  
\zfl{mfig3}
\end{figure}

\begin{figure*}[t]
  \centering
  {\includegraphics[width=0.98\textwidth]{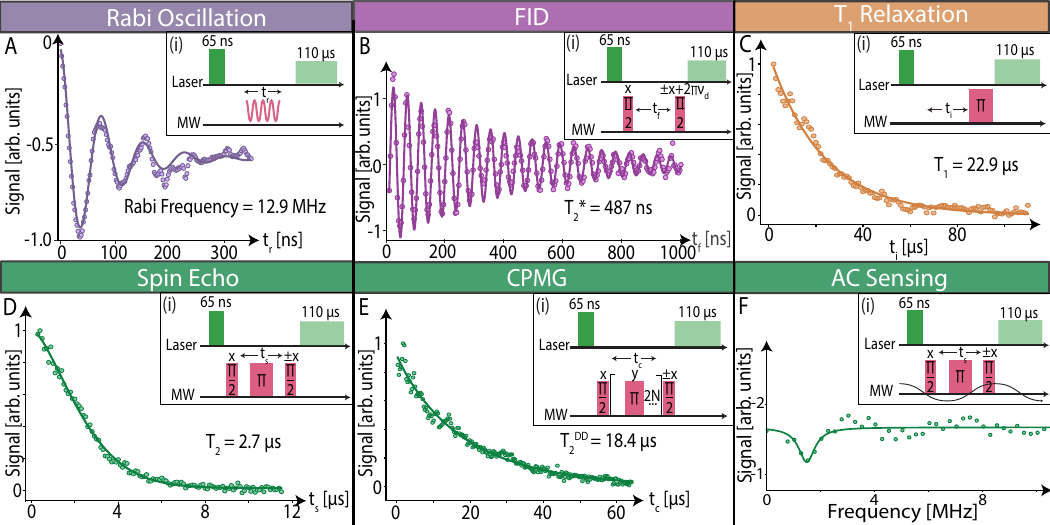}}
  \caption{\T{Optically-detected coherent control} of the photoexcited triplet focusing on the $T_{xy}$ transition. (A) \I{Rabi oscillations} at 12.9 MHz. \I{Inset: (i)} Protocol used. Points are experimental data; solid line is a decaying sinusoidal fit. (B) Ramsey measurement results in $T_2^{\ast}=487{\pm}2$ ns. Second pulse is applied with a 20 MHz ($\nu_d$) frequency offset (\I{Inset} (i)). (C) $T_1$ relaxation measured via inversion recovery, yielding a monoexponential decay with $T_1=22.9{\pm}0.4\mu$s. (C) Spin echo yields $T_2=2.7{\pm}0.1\mu$s. Solid line is a stretched exponential fit. (E) Multipulse CPMG dynamical decoupling with interpulse spacing 148 ns, yielding a monoexponential decay $T_2^{\R{DD}}=18.4{\pm}0.5\mu$s. For all relaxation measurements, each time-point signal was subtracted from the reference signal for background suppression. For Rabi and $T_1$ experiments, the reference signal for each time point repeated the same sequence without MW. For spin-echo and CPMG experiments, the reference signal repeated the same sequence except with a $\pi$-phase added in the $\pi$/2 pulse before the 110$\mu$s readout~\cite{PhysRevB.101.134110}. (F) AC-field sensing using a spin-echo sequence, bias field $B_{x}=6.2$mT, and 7.53$\mu$T signal.}
\zfl{mfig4}
\end{figure*}

\section{ODMR of the photoexcited electronic spin-triplets} 
Using this strategy, \zfr{mfig2}A shows the fluorescence-measured zero field ($B_0{=}0$ mT) ODMR spectrum of the triplet electrons. In the simplest case (\zfr{mfig2}A(i)), the sample is continuously illuminated with 532 nm $\I{cw}$-laser (105 mW), and PL is measured with an avalanche photodiode (APD) via a lock-in amplifier referenced to an applied 1.8 kHz amplitude-modulated MW signal (see SI) ~\cite{SI}. \zfr{mfig2}A displays three prominent resonances corresponding to distinct triplet transitions. The $T_{yz}$ transition exhibits an inverted ODMR contrast, likely from a slightly higher $\ket{T_z}$ steady-state population relative to $\ket{T_y}$ (see \zfr{mfig1}G). 

The entire ODMR scan in \zfr{mfig2}A is completed in ${<}$1 min, reflecting the high SNR in the measurements. Furthermore, the radio-frequency (RF) nature of the $T_{xy}$ transition (${\app}$108 MHz) at ZF is advantageous for practical applications, due to cost-effective, high power amplifiers for these frequencies. We estimate a ${<}$3.4\% (${<}$11 \% for a 10X objective) photon collection efficiency (see SI), indicating substantial potential for future SNR improvement (see Discussion and Outlook). Exclusive $\I{cw}$-illumination mirrors the simplicity of typical quantum sensing with NV centers, but comes at the cost of reduced contrast (cf. \zfr{mfig3}B).

Two significant features emerge from ODMR lineshape analysis of \zfr{mfig2}A. First, relatively narrow linewidths, $\ell{\app}7$ MHz, considering the relatively high density of pentacene molecules in the sample ($~\sim0.1\%$). In \zfr{mfig2}B, the true MW-unbroadened ODMR linewidth is determined to be $\ell_0{=}1.9{\pm}0.1$ MHz from measuring $\ell$ as a function of RF power $P_{\mw}$ and extrapolating backwards using the saturation function $\ell (P_{\mw}) {=} \ell_0 + a\sqrt{P_{\mw}}$). 

Extremely low concentration of paramagnetic defects in the host lattice produces narrow linewidths. In contrast, each NV center is typically associated with ${>}$10-times as many P1 centers~\cite{Belthangady13}, and NV-P1 interactions can significantly broaden ODMR linewidths.

Second, ODMR lineshapes are notably asymmetric, see zoomed low power ($P_{RF}{=}1.45 $mW) spectra in \zfr{mfig2}C. Clear steps in the lineshape can be attributed to the hyperfine interaction with $\Hs$ nuclei in the pentacene molecule. This asymmetry has been noted previously in EPR studies~\cite{yago2007pulsed} and is measured here optically for the first time. These features are more pronounced in the lineshape derivative (\zfr{mfig2}D), allowing us to identify coupling to at least three pairs of $\Hs$ nuclei. The combination of narrow linewidths, asymmetric lineshapes, and a steep derivative renders the PDP system attractive for DC magnetometry~\cite{Barry20, wu2022enhanced}. 

ODMR contrast depends on excitation beam polarization 
for the $T_{xy}$ transition (identical results are found for other transitions). \zfr{mfig2}E displays a pronounced ${\app}$27\% dependence on polarization angle $\xt$. Here $\xt{=}0^{\circ}$ corresponds to polarization along $\zhat$, optimal matching to the excited state transition dipole. As shown in \zfr{mfig2}F, high power $\I{cw}$ RF illumination improves ODMR contrast. Contrast can be significantly improved via pulsed optical illumination, as detailed below (cf. \zfr{mfig3}B).

Finite magnetic field, $B_0$, shifts the ODMR spectra in \zfr{mfig2}A, as anticipated. \zfr{mfig2}G shows the case of $\T{B}_0{=}8$mT along $\xhat$. The shift is visible from the vertical dashed lines referencing \zfr{mfig2}A. Additional broadening can be ascribed to hyperfine coupling within the pentacene molecule and slight misalignment with $\xhat$~\cite{kohler1999magnetic}. Points in \zfr{mfig2}H show measured transition frequencies as a function of $B_0{\parallel}\xhat$ (see SI~\cite{SI}). Calculated transition frequencies (solid lines), obtained
 from diagonalizing $\mH_{\R{sys}}$, show excellent agreement.

The DC sensitivity can be evaluated using the expression: $\eta_{\R{DC}}{=}\sigma \sqrt{\tau}/{\frac{dS}{dF}\gamma_{ij}}~$\cite{Schoenfeld11}, where $\xg_{ij}{=} 9$ MHz/mT is the gyromagnetic ratio of the transition $T_{ij}$ (See S.I), $\frac{dS}{dF}$, the maximal spectral slope, $\sigma$, the noise floor evaluated far-off-resonance and integration time $\tau$ chosen as the settling time of the low-pass lock-in amplifier filter. With a 8 mT bias field (in the linear sensing regime), we obtain for $T_{xy}$ (\zfr{mfig2}H), $\eta_{\R{DC}}{=}2.1$ $\mu$T/$\sqrt{\R{Hz}}$ (volume normalized sensitivity $\eta_{\R{DC}}^{V}{=}35\mu$T $\mu$m$^{3/2}$ Hz$^{-1/2}$) and for $T_{xz}$ (\zfr{mfig2}I),  $\eta_{\R{DC}}{=}1.8$ $\mu$T/$\sqrt{\R{Hz}}$ (volume normalized sensitivity $\eta_{\R{DC}}^{V}{=}30 \mu$T $\mu$m$^{3/2}$ Hz$^{-1/2}$), despite low ODMR contrast (cf. \zfr{mfig2}F) the signal-to-noise-ratio (SNR) is large from high pentacene doping (0.1\%).

\section{Enhanced ODMR contrast via pulsed illumination}
 \zfr{mfig1}E indicates that pulsed optical illumination is more efficient at initialization than cw-illumination, which was implemented with the sequence in \zfr{mfig3}A(i). A laser pulse for period $\qt_p$ was followed by a MW $\pi$-pulse applied to the $T_{xy}$ transition and a 110 $\mu$s readout laser pulse. Optical pulses were generated by chopping the $\I{cw}$-beam with an Acousto-Optic Modulator (AOM). ODMR contrast profiles obtained using this sequence at different pulse powers (\zfr{mfig3}A) show an optimal pulse period $\qt_p^{\R{opt}}$ (dashed line) which moves to shorter times with increasing optical power, yielding an increase in ODMR contrast .

To further improve contrast, we vary excitation laser pulse energy (\zfr{mfig3}B). Fluorescence readout is still performed under $\I{cw}$-illumination (150mW) for time $\qt$ with $\qt_p{=}65$ns (\zfr{mfig3}B(i)). Background subtraction is carried out by performing alternate experiments without and with the MW pulse and subtracting the result. High contrast is observed at optimal readout time ($\qt^{\R{opt}}$). The origin of these profiles evident from taking the difference of the traces in \zfr{mfig1}H. 
ODMR contrast increases dramatically at higher pulse energies, yielding 16.8\% contrast at 83.1 $\mu$J (\zfr{mfig3}C). Considering simulations in \zfr{mfig1}H, we anticipate further gains up to 58\%
contrast at even higher energies, since current experiments were limited by APD saturation beyond 90 $\mu$J, necessitating the use of attenuating ND filters (see SI).  This constraint could be circumvented by using fast MEMS optical switches~\cite{Agiltron} in future experiments to further improve contrast (see Discussion). 

\T{\I{Optically-detected coherent control}} -- \zfr{mfig4} demonstrates coherent electronic control of the photoexcited electrons. Following the \zfr{mfig4}A(i) protocol, \zfr{mfig4}A displays optically detected 12.9 MHz Rabi oscillation for the $T_{xy}$ transition. Driving  coil RF inhomogeneity contributes to the $T_{2\xr} {=}135$ ns Rabi decay. 

\zfr{mfig4}B-D examines the electronic spin coherence properties in the system. Following sequences depicted in the insets, we measure $T_2^{\ast}$ using a Ramsey scheme, $T_1$ through an inversion recovery scheme, and $T_2$ using the Hahn echo sequence, yielding $T_2^{\ast}{=}487{\pm} 2$ ns, $T_1{=}22.9{\pm} 0.4 \mu$s, and $T_2{=}2.7{\pm}0.1 \mu$s, respectively. In the Ramsey measurements (\zfr{mfig4}B), a 20 MHz offset in the second pulse clearly distinguishes the dephasing decay. Additionally, \zfr{mfig4}E shows that employing multipulse dynamical decoupling trains can effectively prolong the $T_2$ coherence time yielding $T_2^{\R{DD}}{=}18.4 {\pm }0.5 \mu$s with a CPMG train spacing of 148 ns. $T_2^{\R{DD}}$ can approach $T_1$ here since the latter is dominated by triplet-ground relaxation rather than spin-lattice relaxation~\cite{Wu19}. Moving the population into the long-lived $\ket{T_z}$ state might allow us to overcome this limit and access longer coherence times (see Discussion). 

 \zfr{mfig4}F shows AC sensing using the $T_{xy}$ transition with a 6.2 mT biased $B_x$ field and a test signal of 7.53 $\mu$T. The obtained sensitivity is $\eta_{\R{AC}}{=}$ 405 nT/$\sqrt{\text{Hz}}$, and the volume-normalized sensitivity is $\eta^V_{\R{AC}}$ = 6.8 $\mu$T $\mu$m$^{3/2}$ Hz$^{-1/2}$ (see SI). We estimate the shot-noise limited pulsed sensitivity, $\eta_\text{pulsed}{\approx} \frac{8}{3\sqrt{3}}\frac{\hbar}{g_e\mu_B}\frac{1}{C_\text{pulsed}\sqrt{\mathscr{N}}}\frac{\sqrt{t_I+T_2+t_R}}{T_2}$~\cite{Barry20}, where $t_I{\app}400 \mu$s is the initialization time (including dead-time for population reset), $t_R{=}110 \mu$s is the readout window (\zfr{mfig3}B), for which the time-averaged contrast is $C_\text{pulsed}{=} 0.117$. $\mathscr{N}$ is the photons collected per measurement and is estimated from count rate 1.9 $\times$ 10$^{12}$ counts/s (PL power of 0.6 $\mu$W at APD). This yields sensitivity estimates of $\eta_{\R{DC}}{=}0.24 \, \text{nT} \, \text{Hz}^{-1/2}$ for the Ramsey sequence and $\eta_{\R{AC}}{=} 43 \, \text{pT} \, \text{Hz}^{-1/2}$ and $6.4 \, \text{pT} \,\text{Hz}^{-1/2}$ for Hahn echo and CPMG sequences, respectively. The corresponding volume-normalized values are: Ramsey $\eta^V_{\R{DC}}{=}18 \, \text{nT} \, \mu\text{m}^{3/2} \, \text{Hz}^{-1/2}$, and $\eta^V_{\R{AC}}{=}3.3 \, \text{nT} \, \mu\text{m}^{3/2} \, \text{Hz}^{-1/2}$ and $486 \, \text{pT} \, \mu\text{m}^{3/2} \, \text{Hz}^{-1/2}$ for Hahn echo and CPMG, respectively. 

\section{Discussion and Outlook}
While calculated sensitivities are idealistic to achieve, since we assumed sensitivity remains unchanged when estimated using a biased magnetic field, they demonstrate scope for improvement in experimentally measured AC and DC sensitivities. We anticipate that the following improvements will enhance sensitivity.

\I{(i) Pulsed laser use:} Due to detector saturation, we used a low-energy initializing laser pulse in this setup, yielding a contrast of only 0.6$\%$. Resolving the detector saturation issue and using a pulsed laser to achieve contrasts of 16.8$\%$ would directly enable an AC sensitivity of 14 nT/$\sqrt{\text{Hz}}$, twice as good as the result reported in Zhou et al.\cite{Zhou20}. Maximum theoretical contrast is estimated to be 58$\%$. \I{(ii) Contrast/photon count optimization:} Pushing more population into the triplet state reduces photon emission, requiring sensitivity optimization based on both total photon counts and ODMR contrast. \I{(iii) Sensing with the $\ket{T_{z}}$ state:} Currently, sensing is done using the $\ket{T_{x}}$ state, so its room temperature lifetime, 35 $\mu$s, sets the upper bound for coherence. The $\ket{T_{z}}$ lifetime is longest 500 $\mu$s. \I{(iv) Deuteration:} ODMR line broadening under a magnetic field can potentially be mitigated by using spin-less or lower-spin magnetic moment nuclei in the environment, such as replacing some protons in the crystal with deuterium. Combining deuteration with low temperatures of 1.5 K resulted in a triplet state linewidth of 120 kHz~\cite{kohler1995single}. \I{(v) Detection:} Optimizing detection for pulsed laser use, refining quantum control, exploiting the long-lived $\ket{T_z}$ state, reducing initialization and readout times, and increasing photon collection efficiency can further improve sensitivity~\cite{wolf2015subpicotesla}(see SI).

Salient features of the chromophore system are here contrasted with conventional quantum sensors. Pentacene molecules in the host matrix act similarly to NV centers, but with higher quantum efficiency (62.5\%~\cite{Takeda02} vs. 45\%~\cite{Suter17}). Since they are not associated with defect centers, every pentacene molecule can function as a quantum sensor. No additional paramagnetic impurities act as decoherence sources (for example in diamond, P1 centers~\cite{bauch2020decoherence}), enabling quantum sensing at significantly higher electronic densities, orders of magnitude higher than NV centers. Even at ${\app}0.1\%$ doping, electronic linewidths remain narrow ($<$7MHz), well-suited for bulk quantum sensing~\cite{Barry20}.  Crystals can be grown to large sizes at low costs, expanding material options for bulk magnetometry~\cite{Barry20} and addressing the material costs and supply limitations of CVD/HPHT diamond manufacturing ~\cite{gicquel2001cvd,pelucchi2022potential}. 


Our work suggests exciting prospects for chemical-based quantum sensing, expanding the design space and applicability of q-sensors. Pentacene is indicative of a wider class of systems with optically-addressable triplet electrons~\cite{tait2015triplet,richert2017constructive,maylander2022accessing, hamachi2021porphyrins,sakamoto2023polarizing} and wherein Hamiltonian parameters like zero-field splitting can be finely tuned through chemical means~\cite{amdur2022chemical}. Synthetic chemistry enables the creation of sensor arrays with precise control over spacing, topology, and concentration. One can envision molecular quantum sensor tags that integrate into various chemical and biological systems as nanoscale fluorescent magnetic field reporters~\cite{Kucsko13,Wu22}. PDP shows potential for high-sensitivity pressure~\cite{cai2014hybrid,macquarrie2013mechanical} and temperature~\cite{Acosta10} sensing due to its deformable lattice and lower melting point, respectively. 

Finally, the ability to hyperpolarize nuclear spins in pentacene-like systems presents opportunities~\cite{eichhorn2014dynamic,Eills23}. $\Hs$ nuclear spins have been polarized to ${>}$30\% levels via relatively simple techniques~\cite{tateishi2014room,eichhorn2014proton}. A key feature is their prolonged nuclear spin-lattice relaxation time, $T_1$. Ref.~\cite{eichhorn2014proton,quan2019polarization} recorded a $\Hs$ $T_1$ exceeding 100hr at 100K and 300mT in pentacene, attributed to the diamagnetic singlet ground state shielding nuclear spins from electronic relaxation. This contrasts with diamond, where $\Cs$ nuclear spins exhibit short $T_1{\sim}$ 5.43 minutes at 100K and 27 mT due to interaction with the paramagnetic NV center~\cite{Jaskula19,beatrez2023electron}. Exploiting long-lived nuclear spins opens new perspectives for gyroscopes~\cite{Ledbetter12,Ajoy12g,Jarmola21}, AC magnetometers~\cite{Sahin21}, and fundamental condensed matter physics research with driven, hyperpolarized nuclei~\cite{beatrez2023critical,Harkins23}.
\begin{acknowledgments}
We gratefully acknowledge discussions with D. Suter, M. Parashar, P. Hautle, Y. Quan, J. Steiner, figure contributions by A. Singh, and funding from AFOSR (FA9550-23-1-0106), DOE BES (DE-SC0020635), NSF TAQS, DNN NNSA (FY24-LB-PD3Ta-P38), EPSRC (EP/S000798/2) and the Royal Society (URF/R1/191297). \I{Note added:} -- When this paper was in final stages, we became aware of Ref.~\cite{Mena24} with complementary results.  
\end{acknowledgments}
\vspace{-5mm} 
%
\vspace{-1mm}

\end{document}